\documentclass[submission,copyright,creativecommons]{eptcs}
\providecommand{\event}{DS Festschrift 2013} 
\usepackage{breakurl}             
\usepackage[latin1]{inputenc}
\usepackage{graphicx}
\usepackage{amssymb} 

\title{Online partial evaluation of sheet-defined functions\\ 
}
\author{Peter Sestoft
\institute{IT University of Copenhagen}
\email{sestoft@itu.dk}}
\def\titlerunning{Partial evaluation of sheet-defined functions}
\def\authorrunning{P. Sestoft}
\begin{document}

\newtheorem{example-thm}{Example}[section]
\newenvironment{example}{\vspace{0.2ex}\begin{example-thm}\rm}%
{\hfill$\Box$\end{example-thm}}

\newcommand{\bibindex}[1]{}  

\maketitle

\begin{abstract}
  We present a spreadsheet implementation, extended with sheet-defined
  functions, that allows users to define functions using only standard
  spreadsheet concepts such as cells, formulas and references,
  requiring no new syntax.  This implements an idea proposed by
  Peyton-Jones and others \cite{PeytonJones:2003:AUserCentred}.

  As the main contribution of this paper, we then show how to add an
  online partial evaluator for such sheet-defined functions.  The
  result is a higher-order functional language that is dynamically
  typed, in keeping with spreadsheet traditions, and an interactive
  platform for function definition and function specialization.

  We describe an implementation of these ideas, present some
  performance data from microbenchmarks, and outline desirable
  improvements and extensions.
\end{abstract}

\section{Introduction}

Spreadsheet programs such as Microsoft Excel, OpenOffice Calc and
Gnumeric provide a simple, powerful and easily mastered end-user
programming platform for mostly-numeric computation.  Yet as observed
by several authors
\cite{Nunez:2000:AnExtended,PeytonJones:2003:AUserCentred},
spreadsheets lack the most basic abstraction mechanism: The ability to
create a named function directly from spreadsheet formulas.

Although most spreadsheet programs allow function definitions in
external languages such as VBA, Java or Python, those languages
have a completely different programming model that many competent
spreadsheet users find impossible to master.  Moreover, the external
language bindings are sometimes surprisingly inefficient.

Here we present an implementation of \emph{sheet-defined functions} in
the style Nuñez \cite{Nunez:2000:AnExtended} and Peyton-Jones et
al. \cite{PeytonJones:2003:AUserCentred} that (1) uses only standard
spreadsheet concepts and notations, no external languages, so it
should be understandable to competent spreadsheet users, and (2) is
very efficient, so that user-defined functions can be as fast as
built-in ones.  The ability to define functions directly from
spreadsheet formulas should (3) encourage the development of shared
function libraries, which in turn should (4) improve reuse,
reliability and upgradability of spreadsheet models.  

Moreover, we believe that \emph{partial evaluation}, or automatic
program specialization, is a plausible tool in the declarative and
interactive context of spreadsheets.  In particular, the
specialization of a function closure \texttt{fv} is achieved through a
simple function call \texttt{SPECIALIZE(fv)}, which creates a new
specialized function and returns a closure for it.  The resulting
specialized function can be invoked immediately and is used exactly
like the given unspecialized one; but (hopefully) executes faster.  No
external files are created and no compilers need to be invoked.  The
design and implementation of this idea are the main contributions of
this paper, presented in sections~\ref{sec-partial-evaluation}
through~\ref{sec-partial-evaluation-examples}.

Our motivation is pragmatic.  A sizable minority of spreadsheet users,
including biologists, physicists and financial analysts, build very
complex spreadsheet models.  This is because it is convenient to
experiment with both models and data, and because the models are easy
to share and distribute.  We believe that one can advance the state of
the art by giving spreadsheet users better tools, rather than telling
them that they should have used Matlab, Java, Python or Haskell
instead.

Another paper, describing complementary aspects of this work, in
particular details of evaluation conditions as well as a case study on
implementing financial functions, appears elsewhere
\cite{Sestoft:2013:SheetDefined};
section~\ref{sec-sheet-defined-functions} of the present paper is
reproduced from that paper.  Many more technical details are given in
a draft technical report \cite{Sestoft:2012:SpreadsheetTechnology}.
Our prototype source code is available from
http://www.itu.dk/people/sestoft/funcalc/.

\section{Sheet-Defined Functions}
\label{sec-sheet-defined-functions}

We first give two motivating examples of sheet-defined functions.

\begin{figure*}[hbt]
  \centering
  \includegraphics[angle=-90,width=0.65\textwidth]{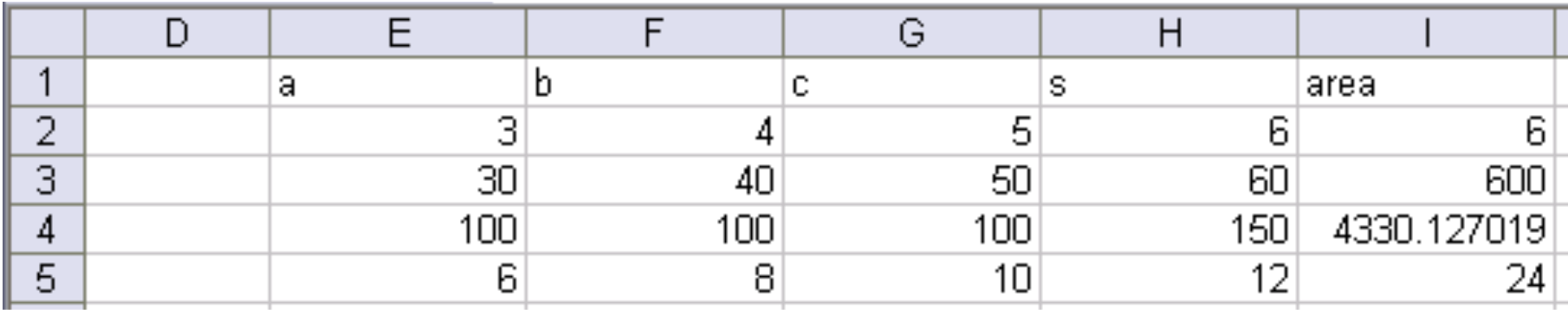}  
  \caption{Triangle side lengths and computed areas, with intermediate
    results in column H\@.}
  \label{fig-funcalc-manual-triangles-excel}
\end{figure*}

\begin{figure*}[hbt]
  \centering
  \vspace{-0.5cm}
  \includegraphics[width=0.72\textwidth]{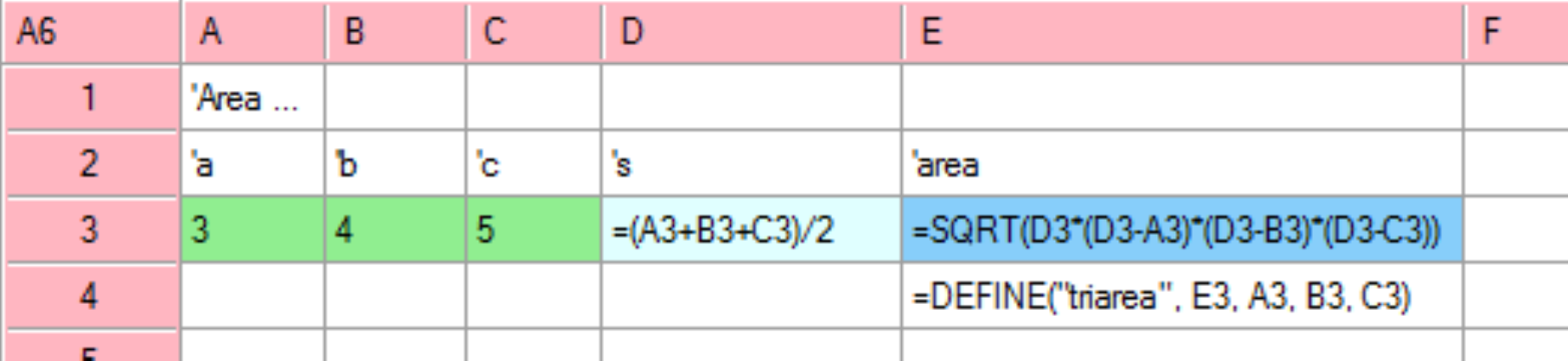}  

  \caption{Function sheet, where \texttt{DEFINE} in E4 creates
    function \texttt{TRIAREA} with input cells A3, B3 and C3, output
    cell E3, and intermediate cell D3\@.}
  \label{fig-funcalc-manual-triarea-sdf}
\end{figure*}

\begin{figure*}[hbt]
  \centering
  \vspace{-0.5cm}
  \includegraphics[width=0.7\textwidth]{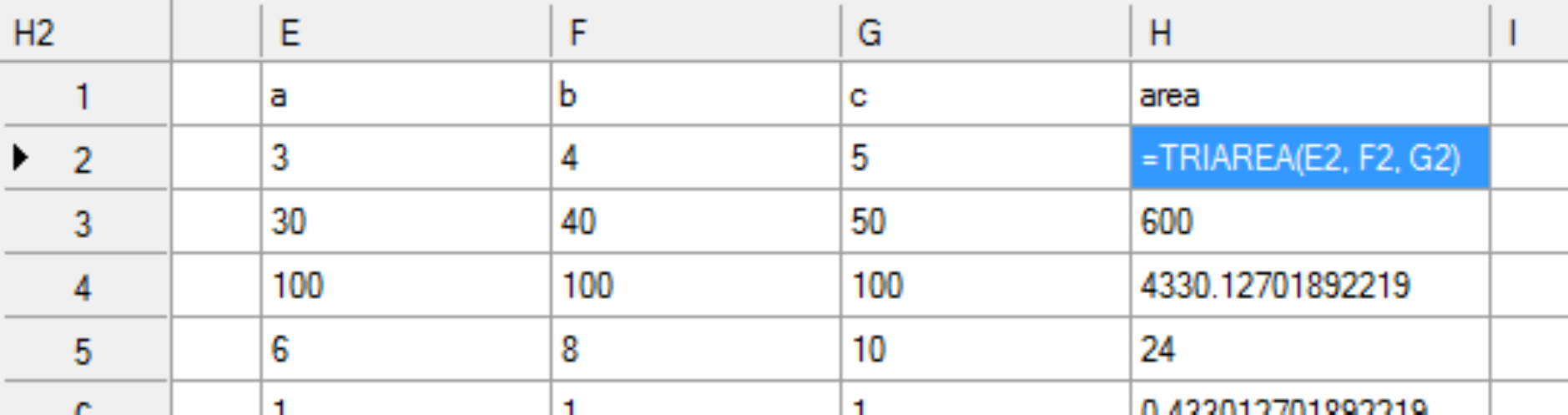}

  \caption{Ordinary
    sheet calling \texttt{TRIAREA}, defined in
    Figure~\ref{fig-funcalc-manual-triarea-sdf}, from cells H2:H5\@.} 
  \label{fig-funcalc-manual-triarea-call}
\end{figure*}

\begin{example}
  Consider the problem of calculating the area of each of a large
  number of triangles whose side lengths $a$, $b$ and $c$ are given in
  columns E, F and G of a spreadsheet, as in
  Figure~\ref{fig-funcalc-manual-triangles-excel}.  The area can be
  computed by the formula $\sqrt{s(s-a)(s-b)(s-c)}$ where $s =
  (a+b+c)/2$ is half the perimeter.  Now, either one must allocate a
  column H to hold the value $s$ and compute the area in column I, or
  one must inline $s$ four times in the area formula.  The former
  pollutes the spreadsheet with intermediate results, whereas the
  latter would create a long expression that is nearly impossible to
  enter without mistakes.  It is clear that many realistic problems
  would require even more space for intermediate results and even more
  unwieldy formulas.

Here we propose instead to define a function, \texttt{TRIAREA} say,
using standard spreadsheet cells and formulas, but on a separate
\emph{function sheet}, and then call this function as needed from the
sheet containing the triangle data.

Figure~\ref{fig-funcalc-manual-triarea-sdf} shows a
function sheet containing a definition of function \texttt{TRIAREA},
with inputs $a$, $b$ and $c$ in cells A3, B3 and C3, the intermediate
result $s$ in cell D3, and the output in cell E3\@.

Figure~\ref{fig-funcalc-manual-triarea-call} shows an ordinary sheet
with triangle side lengths in columns E, F and G, and function calls
\texttt{=TRIAREA(E2,F2,G2)} in column H to compute the triangles'
areas.  There are no intermediate results; these exist only on the
function sheet.  As usual in spreadsheets, it suffices to enter the
function call once in cell H2 and then copy it down column H with
automatic adjustment of cell references.
\end{example}

\newpage

\begin{example}\label{ex-sdf-rept4}
  Microsoft Excel has a built-in function \texttt{REPT($s$,$n$)} that
  computes $s^n$, the string consisting of $n\geq 0$ concatenated
  copies of string $s$.  This built-in function can be implemented
  efficiently as a recursive sheet-defined function
  \texttt{REPT4(s,n)} as shown in figure~\ref{fig-funcalc-rept4-sdf}.
  This implementation is optimal, using $O(\log n)$ string
  concatenation operations, written (\texttt{\&}), for a
  total running time of $O(n \cdot |s|)$, where $|s|$ is the length of
  $s$.
\begin{figure*}[hbt]
  \centering
  \includegraphics[width=0.7\textwidth]{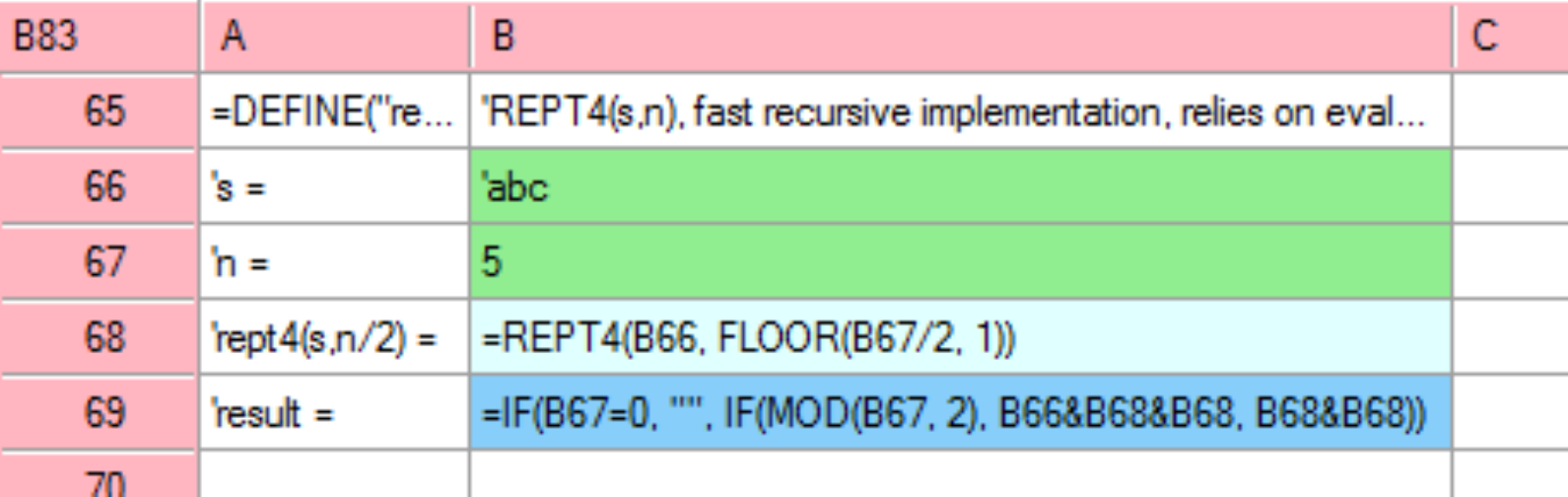}  
  \caption{Recursive function \texttt{REPT4(s,n)} computes $s^n$, the
    concatenation of $n$ copies of string $s$.  Cells B66 and B67 are
    input cells, B69 is the output cell, and B68 holds an intermediate
    result.}
  \label{fig-funcalc-rept4-sdf}
\end{figure*}
\end{example}

In contrast to Peyton-Jones et al., we allow sheet-defined functions
to be \emph{recursively defined}, because that is necessary to
reimplement many built-in functions, and \emph{higher-order}, because
that generalizes many of Excel's ad hoc functions (such as
\texttt{SUMIF}, \texttt{COUNTIF}) and features (such as Goal Seek)\@.

\section{Dual implementation: Interpretation and compilation}

Our spreadsheet implementation, called Funcalc, is written in C\# and
consists of an \emph{interpretive implementation} of ordinary sheets,
described in this section, and a \emph{compiled implementation} of
sheet-defined functions, described in
section~\ref{sec-compiled-implementation}.  The rationale for having
both is that ordinary sheets will be edited frequently but each cell's
formula is evaluated only once every recalculation, whereas a
sheet-defined function will be edited rarely but may be called
thousands of times in every recalculation.  Hence compilation would cause
much overhead and provide little benefit in ordinary sheets, but
causes little overhead and provides much benefit in function sheets.

Funcalc uses the standard notions of workbook, worksheet, cell,
formula and built-in function.  A cell may contain a constant or a
formula (or nothing); a formula in a cell consists of an expression
and a cache for the expression's value.

\subsection{Interpretive implementation}
\label{sec-interpretive-implementation}

The interpretive implementation is used to evaluate formulas on
ordinary sheets.  Spreadsheet formulas are dynamically typed, so
runtime values are tagged by runtime types such as Number, Text,
Error, Array, and Function, with common supertype Value\@.  Such
tagged runtime values will generally be allocated as objects in the
garbage-collected heap.

A formula expression \texttt{e} in a given cell on a given worksheet
is evaluated interpretively by calling \texttt{e.Eval(sheet,col,row)},
which returns a Value object.  A simple interpretive evaluator guided
by the expression abstract syntax tree will involve wrapping of result
values upon return, followed by unwrapping when the value is used.
This overhead is especially noticeable when a 64-bit floating-point
number (type double, which could be stack-allocated) gets wrapped as a
heap-allocated Number object.  An important goal of the compiled
implementation (section~\ref{sec-compiled-implementation}) is to avoid
this overhead.

A technical report \cite{Sestoft:2012:SpreadsheetTechnology} gives
more details about the interpretive implementation and its design
tradeoffs.

\section{Compiled implementation of sheet-defined functions}
\label{sec-compiled-implementation}

The compiled implementation is used to execute sheet-defined
functions.

\subsection{Defining functions and creating closures}
\label{sec-defining-functions}

We add just three new built-in functions to support the definition and
use of sheet-defined functions, including higher-order ones.  There is
a function to define a new sheet-defined function:

\begin{itemize}
\item \texttt{DEFINE("name", out, in1..inN)} creates a function with
  the given name, result cell \texttt{out}, and input cells
  \texttt{in1..inN}, where \texttt{N >= 0}, as shown in
  figure~\ref{fig-funcalc-manual-triarea-sdf} cell E4\@.
\end{itemize}

\noindent
Two other functions are used to create a function value (closure) and
to apply it, respectively:

\begin{itemize}
\item \texttt{CLOSURE("name", e1..eM)} evaluates argument expressions
  \texttt{e1..eM} to values \texttt{a1..aM} and returns a closure for
  the sheet-defined function \texttt{"name"}.  An argument \texttt{ai}
  that is an ordinary value gets stored in the closure, whereas an
  argument that is \texttt{\#NA}, for ``not available'', signifies
  that this argument must be provided later when calling the closure.
  
\item \texttt{APPLY(fv, e1..eN)} evaluates \texttt{fv} to a closure,
  evaluates argument expressions \texttt{e1..eN} to values
  \texttt{b1..bN}, and applies the closure by using the \texttt{bj}
  values for those arguments in the closure that were \texttt{\#NA} at
  closure creation.
\end{itemize}

\noindent 
The \texttt{\#NA} mechanism provides a very flexible way to create
closures, or partially applied functions.  For instance, given a
function \texttt{MONTHLEN(y,m)} to compute the number of days in month
\texttt{m} of year \texttt{y}, we may create the closure
\texttt{CLOSURE("MONTHLEN", 2013, \#NA)} which computes the number of
days in any month in 2013, or we may create the closure
\texttt{CLOSURE("MONTHLEN", \#NA, 2)} which computes the number of
days in February in any year.  This notational flexibility is rather
unusual from a programming language perspective, but fits well with
the standard spreadsheet usage of \texttt{\#NA} to denote a value that
is not (yet) available.

The built-in functions \texttt{DEFINE}, \texttt{CLOSURE} and
\texttt{APPLY} provide the general computation model for higher-order
functions in our implementation, and are the only means needed to
define functions, create closures, and call closures.  See also
figure~\ref{fig-clo-spec-apply-bench}.  All the required syntax is
already present in standard spreadsheet implementations.

Section~\ref{sec-partial-evaluation} shows that these functions fit
excellently with the single new function \texttt{SPECIALIZE} needed to
partially evaluate closures.

\subsection{Overall compilation process}
\label{sec-compilation-process}

The compiled implementation of sheet-defined functions generates CLI
(.NET) bytecode at runtime.  Runtime bytecode generation maintains the
interactive style of spreadsheet development and some portability,
while achieving good performance because the .NET just-in-time
compiler turns the bytecode into efficient machine code.

The compilation of a sheet-defined function proceeds in these steps:

\begin{enumerate}
\item\label{item-compile-dependency-graph} Build a dependency graph of
  the cells transitively reachable, by cell references, from the
  output cell.
  
\item\label{item-compile-topological} Perform a topological sort of
  the dependency graph, so a cell is preceded by all cells that it
  references.  It is illegal for a sheet-defined function to have
  static cyclic dependencies.
  
\item\label{item-compile-inline} If a cell in the graph is referred
  only once (statically), inline its formula at its unique
  occurrence.  This saves a local variable at no cost in code size or
  computation time.
  
\item\label{item-compile-evaluation-condition} Using the dependency
  graph, determine the evaluation condition
  \cite{Sestoft:2013:SheetDefined} for each cell; build a new
  dependency graph that takes evaluation conditions into account; and
  redo the topological sort.  The result is a list of ComputeCell
  objects, each containing a cell \texttt{c} and an associated
  variable \verb|v_c|.
  
\item\label{item-compile-generate-code} Generate CLI bytecode:
  traverse the list in forward topological order, and for each pair of
  a cell \texttt{c} and associated variable \verb|v_c|, generate code
  for the assignment \verb|v_c = <code for c's formula>|.
\end{enumerate}

\noindent
The evaluation conditions mentioned in
step~\ref{item-compile-evaluation-condition} are needed to correctly
evaluate recursive functions such as that in
example~\ref{ex-sdf-rept4}; see \cite{Sestoft:2013:SheetDefined} for
more details.

\begin{figure}[htbp]
  \centering
  \includegraphics[width=1.0\textwidth]{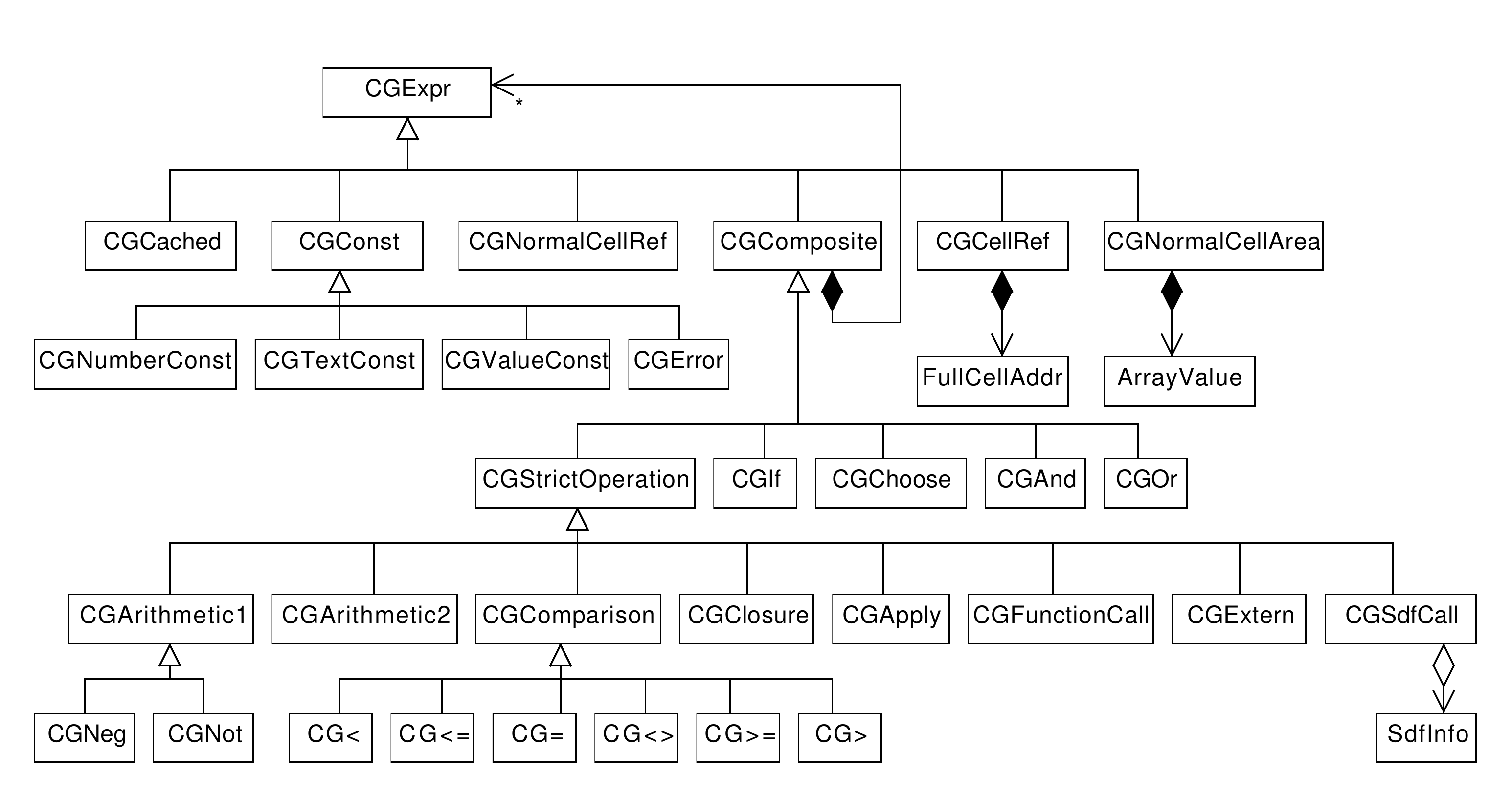}  
  \caption{Class diagram for Funcalc's expression abstract syntax.
    All expressions are side effect free except possibly calls to
    external functions (CGExtern)\@.
    The rather elaborate inheritance structure facilitates ordinary
    code generation as well as partial evaluation.  A CGClosure is a
    \texttt{CLOSURE()} call, CGApply is an \texttt{APPLY()} call,
    CGFunctionCall is a call to any other built-in function, CGExtern
    is a call to an external .NET function, and CGSdfCall is a call to
    a sheet-defined function.  This is used in
    section~\ref{sec-partial-evaluation-process}.} 
  \label{fig-funcalc-cgexpr-classdiagram}
\end{figure}

The formula in a cell \texttt{c} contains an expression represented by
abstract syntax of type CGExpr (for \emph{c}ode \emph{g}enerating
\emph{expr}ession); see figure~\ref{fig-funcalc-cgexpr-classdiagram}.

\subsection{Context-dependent compilation}
\label{sec-compilation-context-dependent}

This section describes the compilation, basically
step~\ref{item-compile-generate-code} in
section~\ref{sec-compilation-process}, of a formula expression
represented by the abstract syntax shown in
figure~\ref{fig-funcalc-cgexpr-classdiagram}.  Compilation to CLI
bytecode is performed at runtime by a straightforward recursive
traversal of the abstract syntax tree
\cite{Sestoft:2012:ProgrammingLanguage}.  The main embellishments of
the code generation serve (1) to avoid wrapping 64-bit floating-point
numbers as heap-allocated Number objects, (2) to represent and
propagate error values efficiently, and (3) to implement proper tail
recursion for sheet-defined functions.  Point (2) is required by
spreadsheet semantics: a failed computation must produce an error
value, and such error values must propagate from operands to results
in any context, such as arithmetics, comparisons, conditions, and
function calls.

Achieving (3) is near-trivial, using the CLI ``\texttt{tail.}''\
instruction prefix (as in
example~\ref{ex-partial-evaluation-expsample}); how to achieve (1) and
(2) is described in the remainder of this section.

\subsubsection{No value wrapping}
\label{sec-compiled-no-wrapping}

The simplest compilation scheme, implemented by the method
\texttt{Compile} on expressions, generates code to emulate
interpretive evaluation.  The call \texttt{e.Compile()} must generate
code that, when executed, leaves the value of \texttt{e} on the stack
top as a Value object.

However, using \texttt{Compile} would wrap every intermediate result
as an object of a subclass of Value, which would be inefficient, in
particular for numeric operations.  In an expression such as
\texttt{A1*B1+C1}, the intermediate result \texttt{A1*B1} would be
wrapped as a Number object, only to be immediately unwrapped before
the addition.  The creation of that Number object would dominate all
other costs, because it requires allocation in the heap and causes
work for the garbage collector.

To avoid runtime wrapping of results that will definitely be used as
numbers, we introduce another compilation method.  The call
\texttt{e.CompileToDoubleOrNan()} must generate code that, when
executed, leaves the value of \texttt{e} on the stack as a 64-bit
floating-point value.  If the result is an error value, then the
number will be a NaN\@.

\subsubsection{Error propagation}
\label{sec-compiled-error-propagation}

When computing with naked 64-bit floating-point values, we represent
an error value as a NaN and use the 51 bit ``payload'' of the NaN to
distinguish error values, as per the \textsc{ieee754} standard \cite[section
6.2]{IEEE:2008:FloatingPoint}.  Since arithmetic operations and
mathematical functions preserve NaN operands, we get error propagation
for free.  For instance, if \texttt{d} is a NaN, then
\texttt{Math.Sqrt(6.1*d+7.5)} will be a NaN with the same payload,
thus representing the same error.  As an alternative to error
propagation via NaNs, one could use CLI exceptions, but these turn out
to be several orders of magnitude slower.  This inefficiency matters
little in well-written C\# software, which should not use exceptions
for control flow, so a single exception would terminate the software.  By
contrast, a spreadsheet workbook may have thousands of cells that have
error values during workbook editing, and these may be
reevaluated whenever a single cell has been edited.

If both \texttt{d1} and \texttt{d2} are NaNs, then by the
\textsc{ieee754} standard \texttt{d1+d2} must be a NaN with the same
payload as one of \texttt{d1} and \texttt{d2}.  Hence at most one
error can be propagated at a time, but that fits well with spreadsheet
development: The user fixes one error, only to see another error
appear, then fixes that one, and so on.

\subsubsection{Compilation of comparisons}
\label{sec-compiled-comparisons}

According to spreadsheet principles, comparisons such as
\texttt{B8>37} must propagate errors, so that if B8 evaluates to an
error, then the comparison evaluates to the same error.  When
compiling a comparison we cannot rely on NaN propagation; a comparison
involving one or more NaNs is either true or false, not undefined, in
CLI \cite[section III.3]{Ecma:2012:CLI}.  Hence we introduce yet
another compilation method on expressions.  

The call \texttt{e.CompileToDoubleProper(ifProper,ifBad)} must
generate code that evaluates \texttt{e}; and if the value is a non-NaN
number, leaves it on the stack top as a 64-bit floating-point value
and continues with the code generated by \texttt{ifProper}; else, it
leaves the value in a special variable and continues with the code
generated by \texttt{ifBad}.

Here \texttt{ifProper} and \texttt{ifBad} are themselves code
generators, which generate the success continuation and the failure
continuation \cite{Strachey:1974:Continuations} of the evaluation of
\texttt{e}.  

The default implementation of \texttt{e.CompileToDoubleProper} uses
\texttt{e.CompileToDouble} to generate code to evaluate \texttt{e} to
a (possibly NaN) floating-point value, then adds code to test at
runtime that the result is not NaN\@.  In some cases, such as
constants, this test can be performed at code generation time, thus
resulting in faster code.

\subsubsection{Compilation of conditions}
\label{sec-compiled-conditions}

Like other expressions, a conditional \texttt{IF(e0,e1,e2)} must
propagate errors from \texttt{e0}, so if \texttt{e0} gives an error
value, then the entire conditional expression gives the same error
value.  For this purpose we introduce a fourth and final compilation
method on expressions.

The call \texttt{e.CompileCondition(ifT,ifF,ifBad)} must generate code
that evaluates \texttt{e}; and if the value is a non-NaN number
different from zero, it continues with the code generated by
\texttt{ifT}; if it is non-NaN and equal to zero, continues with the
code generated by \texttt{ifF}; else, leaves the value in a special
variable and continues with the code generated by \texttt{ifBad}.

For instance, to compile \texttt{IF(e0,e1,e2)}, we compile \texttt{e0}
as a condition whose \texttt{ifT} and \texttt{ifF} continuations
generate code for \texttt{e1} and \texttt{e2}.

\subsubsection{On-the-fly optimizations}
\label{sec-compiled-optimizations}

The four compilation methods provide some opportunities for making
local optimizations on the fly.  For instance, in the comparison
\texttt{B8>37} we must test whether B8 is an error, whereas the
constant 37 clearly need not be tested at runtime.  Indeed, method
\texttt{CompileToDoubleProper} on a NumberConst object performs the
error test at code generation time instead.

As another optimization, to compile the unary logical operator
\texttt{NOT(e0)} as a condition, we simply compile \texttt{e0} as a
condition, swapping the \texttt{ifT} and \texttt{ifF} generators.
This is useful because evaluation conditions (see
step~\ref{item-compile-evaluation-condition} in
section~\ref{sec-compilation-process}) are likely to contain such
``silly'' negations.

\section{Performance}
\label{sec-evaluation-performance}

As a non-trivial micro-benchmark for the code generation scheme
outlined in section~\ref{sec-compiled-implementation}, consider the
cumulative distribution function of the normal distribution $N(0,1)$,
known as \texttt{NORMSDIST} in Excel\@.  It can be implemented as a
sheet-defined function using 20 formulas and 15 floating-point
constants, and in C, C\# and VBA using roughly 30 lines of code.

Figure~\ref{fig-funcalc-normdistcdf-performance} compares the
performance of several equally accurate implementations.  It shows
that a sheet-defined function in our prototype implementation may be
just 2.5 times slower than a function written in a ``real''
programming language such as C or C\#, and considerably faster than
one written in Excel's macro language VBA, and faster than Excel's own
built-in function \texttt{NORMSDIST}.

\begin{figure}[htbp]
  \centering
  \begin{tabular}{l|r}\hline\hline
 Implementation & Time (ns)\\\hline

Sheet-defined function & 118 \\

C\# & 47 \\

C (gcc 4.2.1 -O3) & 51 \\

Excel 2007 VBA function & 1925 \\

Excel 2007 built-in \texttt{NORMSDIST} & 993 \\\hline\hline
  \end{tabular}
  \caption{Running time (ns per call) for the cumulative distribution
    function for the normal distribution. Experimental platform:
    Intel Core 2 Duo 2.66 GHz, Windows 7, .NET 4.5, MacOS X\@.
    Average of 200~000 to 1~000~000 calls; the observed variation between
    repeated experiments is less than 10\%.}
  \label{fig-funcalc-normdistcdf-performance}
\end{figure}

As a more substantial evaluation of the performance of sheet-defined
functions, Sørensen reimplemented many of Excel's built-in financial
functions as sheet-defined functions
\cite{Sestoft:2013:SheetDefined,Soerensen:2012:AnEvaluation}.  The
implementation naturally uses recursive and higher-order functions.
In almost all cases, the sheet-defined functions are faster than
Excel's built-ins; the exceptions are functions that involve search
for the zero of a function, for which a rather naive algorithm was
used in the reimplementation.

\section{Partial evaluation of sheet-defined functions}
\label{sec-partial-evaluation}

As described in section~\ref{sec-defining-functions}, the function
call \texttt{CLOSURE("name", a1, ..., aM)} constructs a function value
\texttt{fv}, or closure, in the form of a partial application of
function \texttt{name}.  The closure \texttt{fv} is just a package of
the underlying named sheet-defined function and some early,
non-\texttt{\#NA}, arguments for it.  Applying it using
\texttt{APPLY(fv, b1, ..., bN)} simply inserts the values of
\texttt{b1}\ldots\texttt{bN} instead of the \texttt{\#NA} arguments
and then calls the underlying sheet-defined function; this is no
faster than calling the original function.

However, if the closure \texttt{fv} is to be called more than once, it
may be worthwhile to perform a \emph{specialization} or \emph{partial
  evaluation} of the underlying sheet-defined function with respect to
the non-\texttt{\#NA} values among the arguments
\texttt{a1}\ldots\texttt{aM}.  In Funcalc, this can be done by the
built-in function \texttt{SPECIALIZE}:

\begin{itemize}
\item \texttt{SPECIALIZE(fv)} takes as argument a closure \texttt{fv},
  generates code for a new function, and returns a closure
  \texttt{spfv} for that function.  The resulting closure can be used
  exactly as the given closure \texttt{fv}; in particular, it can be
  called as \texttt{APPLY(spfv, b1, ..., bN)}, and will produce the
  same result as \texttt{APPLY(fv, b1, ..., bN)}, but possibly faster.

  More precisely, if the given closure \texttt{fv} contains \texttt{N}
  arguments with value \texttt{\#NA}, and so has arity \texttt{N}, then the newly
  created function \texttt{clo\#f} will have arity \texttt{N}, and the
  closure \texttt{spfv} will have form \texttt{CLOSURE("clo\#f",
    \#NA...\#NA)} with \texttt{N} occurrences of \texttt{\#NA}\@.  The
  name \texttt{clo\#f} of the new function is the concatenation of the
  print representation \texttt{clo} of the closure \texttt{fv} and an
  internal function number \texttt{\#f}.
\end{itemize}

\noindent
It follows that \texttt{SPECIALIZE(CLOSURE("name", a1, ..., aM))} will
partially evaluate function \texttt{"name"} with respect to the
non-\texttt{\#NA} arguments among \texttt{a1...aM}, and then will
return an \texttt{APPLY}-callable closure for the specialized
function.

Often, the specialized function is faster than the general one, and
often the specialized function can be generated once and then applied
many times.  For instance, this may be the case when finding a root or
computing the integral of a function, or when doing a Monte Carlo
simulation, in which all parameters except one are fixed.

The remainder of this section briefly introduces partial evaluation
and features of particular interest in a spreadsheet setting.
Section~\ref{sec-partial-evaluation-process} then explains how the
\texttt{SPECIALIZE} function has been implemented, and
section~\ref{sec-partial-evaluation-examples} shows some examples of
its use.

\begin{figure}[htbp]
  \centering
  \includegraphics[width=0.7\textwidth]{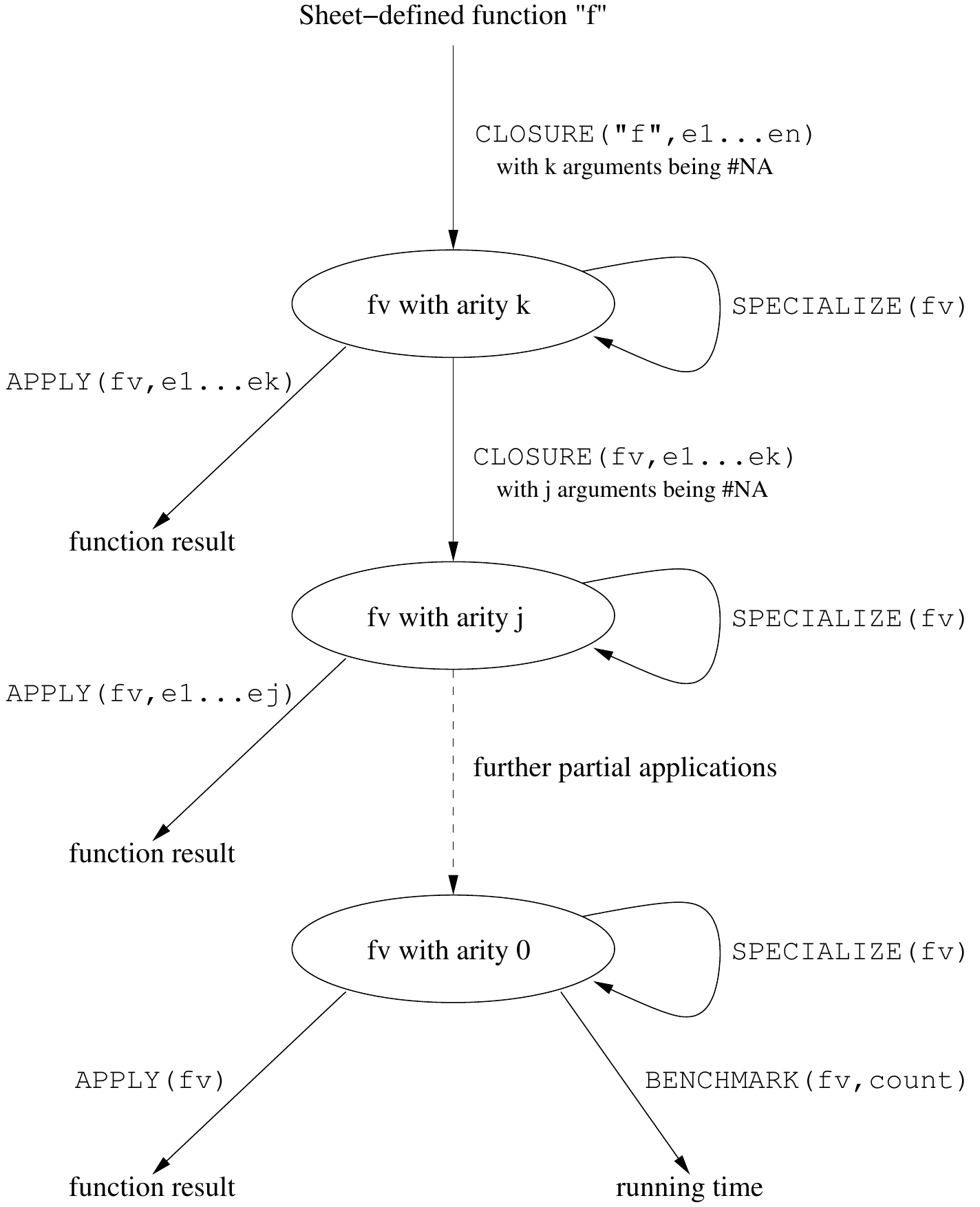}  
  \caption{The relation between functions \texttt{CLOSURE} and
    \texttt{APPLY} for closure creation and closure evaluation
    (section~\ref{sec-defining-functions}), function
    \texttt{SPECIALIZE} for closure specialization
    (section~\ref{sec-partial-evaluation}), and function
    \texttt{BENCHMARK}, for 0-arity closure benchmarking
    (section~\ref{sec-other-features}).}
  \label{fig-clo-spec-apply-bench}
\end{figure}

\subsection{Background on partial evaluation}

Automatic specialization, or partial evaluation, has been studied for
a wide range of languages in many contexts and for many purposes
\cite{Hatcliff:1998:PartialEvaluation,Jones:1993:PartialEvaluation}.
Partial evaluation of a function requires that values for some of the
function's arguments are available.  These are called \emph{static} or
\emph{early} arguments and correspond exactly to the early
(non-\texttt{\#NA}) arguments of a Funcalc function closure; see
section~\ref{sec-defining-functions}.  The remaining \emph{dynamic} or
\emph{late} arguments will be available only when the specialized
function is applied; these correspond to those given as \texttt{\#NA}
when the closure was created.

The result of specializing a function closure with respect to the
given static arguments is a new \emph{residual function} whose
parameters are exactly the dynamic (\texttt{\#NA}) ones from the
closure.  

During partial evaluation, an expression to be specialized may either
be \emph{fully evaluated} to obtain a value, provided all required
variables are static, or it may be \emph{residualized} to obtain a
residual expression, if some required variables are dynamic, or if we
decide to not fully reduce the expression for some other reason.  

Partial evaluation may be performed by \emph{offline} methods, in
which the specialization proper is preceded by a binding-time analysis
that classifies expressions as static or dynamic, or by \emph{online}
methods, where the classification into static and dynamic is performed
during specialization \cite[chapter 7]{Jones:1993:PartialEvaluation}.
Online partial evaluation offers more opportunities for exploiting
available data than offline methods \cite{Ruf:1992:OpportunitiesFor},
whereas offline methods support specializer self-application better,
and may be faster than online methods because tests for a values'
staticness are performed in advance, not repeatedly during
specialization. 

In this work we use online methods, both because we are not concerned
with self-application, and because online methods seem well
aligned with the highly dynamic nature of spreadsheets.

\subsection{Partial evaluation in a spreadsheet context}

Specialization in the context of sheet-defined functions appears to
offer new opportunities, for several reasons:

\begin{itemize}  
\item The declarative computation model makes specialization fairly
  easy.  All values are immutable and all expressions are side effect
  free (except possibly external calls).  The main sources of
  complication are (1) operations that are volatile or that update or
  rely on external state; and (2) recursive function calls.  In both
  respects one can draw on a large body of experience from
  specialization of other dynamically typed languages, notably Scheme;
  see eg.\ Bondorf and Danvy
  \cite{Bondorf:1991:AutomaticAutoprojection:SCP,Bondorf:1991:AutomaticAutoprojection},
  and Ruf, Weise and others
  \cite{Ruf:1993:TopicsIn,Ruf:1992:OpportunitiesFor,Weise:1991:AutomaticOnline}.
  
\item Actual bytecode generation for a specialized function can use
  the same machinery as for non-specialized ones; in particular,
  specialized functions are not penalized by poorer code generation.

\item Specialization of sheet-defined functions that involve volatile
  functions, such as \texttt{RAND} and \texttt{NOW}, requires some
  care.  A volatile function's result should always be considered
  dynamic; the randomness or external state inherent in a volatile
  function is similar to a side effect and hence should be
  residualized as realized already by Bondorf and Danvy
  \cite{Bondorf:1991:AutomaticAutoprojection}.

\item The language of sheet-defined functions is a higher-order
  functional language, and a function can be specialized with respect
  to a function.

\item The spreadsheet environment is interactive and all evaluation
  and specialization takes place in the same environment, containing
  data, the original program and the specialized program.  Any value
  produced during evaluation of a Funcalc spreadsheet can be
  represented by a CGValueConst expression
  (figure~\ref{fig-funcalc-cgexpr-classdiagram}), which is an abstract
  syntax tree node that simply holds a reference to that (immutable)
  value.  Hence there is no need to marshall higher-order
  (function-type) values as data, nor to subsequently restore them as
  function values.  This avoids the problem of cross-stage persistence
  \cite{Neverov:2004:CrossStage}, and the problem of finding which
  functions appear in dynamic context \cite[section
  3.3.5]{Ruf:1993:TopicsIn} when constructing a residual program in
  text form.

\item The seemingly cumbersome split of specialization
  (\texttt{SPECIALIZE}) from closure creation (\texttt{CLOSURE}) means
  that an already-specialized function may be further partially
  applied using \texttt{CLOSURE} and then specialized using
  \texttt{SPECIALIZE}; see
  example~\ref{ex-partial-evaluation-add-multistage} and
  figure~\ref{fig-clo-spec-apply-bench}.  It also means that a
  higher-order library function may specialize an unknown closure
  passed to it as an argument.  Using the timing function
  \texttt{BENCHMARK} (section~\ref{sec-other-features}) it may even
  measure whether the specialized function is faster than the original
  one.
\end{itemize}

\noindent
Automatic specialization of sheet-defined functions permit generality
without performance penalties.  For instance, a company or a research
community can develop general financial or statistical functions, and
rely on automatic specialization to create efficient specialized
versions, removing the need to develop and maintain hand-specialized
ones.

\section{The partial evaluation process}
\label{sec-partial-evaluation-process}

Specialization, or partial evaluation, of a sheet-defined function is
performed through a single additional built-in function
\texttt{SPECIALIZE(fv)} that takes as argument a function closure
\texttt{fv}.  It then generates a specialized version of \texttt{fv}'s
underlying sheet-defined function, based on the early arguments
included in the closure \texttt{fv}.

Partial evaluation of a sheet-defined function works on its
representation as a list of ComputeCell objects, as produced by
steps~\ref{item-compile-dependency-graph}
through~\ref{item-compile-evaluation-condition} in
section~\ref{sec-compilation-process}.  The result is a new
ComputeCell list containing specialized versions of the existing
expressions.  Section~\ref{sec-partial-evaluation-cgexpr-terms}
describes the processing of most expressions, except for function
calls, the source of most complications, which are described in
section~\ref{sec-partial-evaluation-function-calls}.

The specialized ComputeCell list is subsequently used to generate
bytecode for the specialized function, via the compilation functions
described in section~\ref{sec-compilation-context-dependent}.
Furthermore, the ComputeCell list is saved to permit further
specialization of the newly specialized sheet-defined function; this
rather unusual functionality comes for free.

We use classical polyvariant specialization
\cite{Bulyonkov:1984:Polyvariant}, so the residual program consists of
any number of specialized variants of existing sheet-defined
functions.  Such specialized variants may call each other, thus
permitting loops and (mutual) recursion in the residual program.

Each specialized function is given a unique name.  For instance, if
the display value of the given closure \texttt{fv} is
\texttt{ADD(42,\#NA)}, then the result of \texttt{SPECIALIZE(fv)} may
be named \texttt{ADD(42,\#NA)\#117}, where \texttt{\#117} is a unique
internal function number.

The specialized functions will be cached, so that two closures that
are equal (based on underlying function and argument values) will give
rise a single shared specialized function.  That avoids some wasteful
specialization and also is the obvious way to allow for loops (via
recursive function calls) in specialized functions.

The specialization of CGExpr expressions, described in
section~\ref{sec-partial-evaluation-cgexpr-terms} below, takes place
in a partial evaluation environment \texttt{pEnv} that is initialized
and updated as follows.  Initially, \texttt{pEnv} maps the cell
address of each static (non-\texttt{\#NA}) input cell to a constant
representing that input cell's value.  Moreover, \texttt{pEnv} maps
each remaining (dynamic, or \texttt{\#NA}) input cell address to a new
CGCellRef expression representing a residual function argument.

During partial evaluation of a pair \texttt{(ca,e)} in the ComputeCell
list, consisting of a cell address and a expression, the \texttt{pEnv}
is extended as follows.  If the result of partially evaluating the
formula \texttt{e} in cell \texttt{ca} is a CGConst, we extend
\texttt{pEnv} so it maps \texttt{ca} to that constant; then the
constant will be inlined at all subsequent occurrences.  Otherwise,
create a fresh local variable as a copy of the existing cell variable
\texttt{ca}, and add a ComputeCell to the resulting list that will, at
runtime, evaluate the residual expression and store its value in the
new local variable.  Also, extend \texttt{pEnv} to map \texttt{ca} to
the new local variable, so that subsequent references to cell
\texttt{ca} will refer to the new local variable and thereby at
runtime will fetch the value computed by the residual expression.

When the residual ComputeCell list is complete, the dependency graph
is built, a topological sort is performed, use-once cells are inlined,
evaluation conditions are recomputed, and code is generated, following
steps~\ref{item-compile-dependency-graph}
through~\ref{item-compile-generate-code} in
section~\ref{sec-compilation-process} as for a normal sheet-defined
function.

\subsection{Partial evaluation of CGExpr terms}
\label{sec-partial-evaluation-cgexpr-terms}

Partial evaluation of a given cell's formula expression proceeds by
rewriting a given CGExpr term
(figure~\ref{fig-funcalc-cgexpr-classdiagram}) to a residual CGExpr
term, as follows.

\begin{itemize}
\item Partial evaluation of an expression of class CGConst or one of
  its subclasses produces that expression itself.

\item Partial evaluation of a function-sheet cell reference
  CGCellRef(\texttt{c}) produces a CGConst static value if cell
  \texttt{c} is a static input cell or another cell that has been
  reduced to a CGConst subclass; otherwise it produces the given
  expression CGCellRef(\texttt{c}) itself.  This avoids inlining (and
  hence duplication) of residual computations, while still exposing
  static values to further partial evaluation.

\item Partial evaluation of an ordinary-sheet cell reference
  CGNormalCellRef(\texttt{c}) produces that expression itself, not the
  value currently found in the referred ordinary-sheet cell
  \texttt{c}, because that value might change before the residual
  sheet-defined function gets called.  This allows cells on an
  ordinary sheet to be used as ``external parameters'' of a
  specialized function.

\item Partial evaluation of an ordinary-sheet area reference
  CGNormalCellArea(\texttt{area}) produces that expression itself, not
  the values currently found in the referred cells, because those
  values might change before the residual sheet-defined function gets
  called.  

\item Partial evaluation of expressions of class CGStrictOperation and
  most of its subclasses proceeds uniformly as follows.  Partially
  evaluate the argument expressions, and if they are all constants,
  then evaluate the operation as usual; otherwise residualize the
  operation.  In particular, this holds for CGArithmetic1,
  CGArithmetic2, CGComparison and CGFunctionCall, except for volatile
  functions.  A volatile function such as \texttt{NOW()} and
  \texttt{RAND()} should always be residualized, not evaluated, during
  partial evaluation.  For instance, a sheet-defined function might
  perform a stochastic simulation, using the condition
  \texttt{RAND()<0.2} to choose between two scenarios, as in
  example~\ref{ex-sdf-expsample}.  Early evaluation of
  \texttt{RAND()<0.2} would make all executions of the residual
  function choose the same scenario, which would make the specialized
  closure behave differently than the original one.

  The exceptions to the general partial evaluation of
  CGStrictOperation are CGApply and CGSdfCall (residualize to avoid
  infinite loops, see
  section~\ref{sec-partial-evaluation-function-calls}), CGFunctionCall
  (residualize when the called built-in function is volatile), and
  CGExtern (residualize to avoid specialization-time side effects).

\item Partial evaluation of a CGClosure expression follows the general
  CGStrictOperation scheme for partial evaluation.  First it reduces
  its argument expressions.  If all are constant, then it calls the
  interpretive applier corresponding to built-in function
  \texttt{CLOSURE} and produces a CGValueConst that wraps a
  FunctionValue containing the given sheet-defined function and the
  given parameters; otherwise it residualizes.  

\item Partial evaluation of a call to a sheet-defined function
  (CGSdfCall) is discussed separately in
  section~\ref{sec-partial-evaluation-function-calls}.

\item Partial evaluation of a CGApply(\texttt{e0}, \texttt{e1},
  \ldots, \texttt{en}) expression should first reduce all operands,
  both the function expression \texttt{e0} and its arguments.  If the
  function expression in \texttt{e0} is static and is a FunctionValue
  wrapped in a CGValueConst, then partial evaluation can produce a
  CGSdfCall expression, otherwise it must residualize to a CGApply
  based on the residual operand expressions.  Even if both the
  function and all the arguments are static values, it is dangerous to
  actually call the indicated sheet-defined function, as this could
  result in an infinite loop.

  It is worth pondering whether a more aggressive evaluation is
  possible when the function expression \texttt{e0} is static and
  hence is a known FunctionValue\@.  Could we simply further process
  it as if partially evaluating a CGSdfCall expression, using the
  exact same machinery?


\item Partial evaluation of CGIf(\texttt{e0}, \texttt{e1},
  \texttt{e2}) or CGChoose(\texttt{e0}, \texttt{e1}, \ldots,
  \texttt{en}) should produce the result of partially evaluating the
  relevant branch \texttt{ei} if the first expression \texttt{e0} is a
  static value.  Otherwise, they must residualize to a CGIf or
  CGChoose constructed from the residual argument expressions.

\item Partial evaluation of a CGAnd expression, short-cut style, can
  proceed as follows.  Each argument is partially evaluated in turn,
  from left to right.  If the result is constant false (zero), then
  the residual expression for the entire CGAnd is the constant false;
  if the result is constant true (non-zero) then it is ignored; and if
  the result is non-constant, then it is kept for possible inclusion
  in the residual expression.  If no argument reduces to false, then
  the residual expression for the entire CGAnd is the conjunction of
  the residual expressions of the non-true arguments.  In case all
  constant arguments were true, the result is the empty conjunction,
  that is, true.  

\item Partial evaluation of a CGOr expression is dual to CGAnd: just
  swap false and true, and zero and non-zero, in the description
  above.

\item A CGCachedExpr expression may be wrapped around the conditions
  of IF and CHOOSE, for use in evaluation conditions.  Since we ignore
  evaluation conditions during partial evaluation, partial evaluation
  of a CGCachedExpr should simply partially evaluate the enclosed
  expression.
\end{itemize}

\subsection{Partial evaluation of function calls}
\label{sec-partial-evaluation-function-calls}

Partial evaluation of a function call, whether a direct call CGSdfCall
of a named sheet-defined function or a call CGApply of a function
closure via \texttt{APPLY}, poses special challenges that may cause
partial evaluation to fail to terminate.  There are two ways that
partial evaluation may go on indefinitely:

\begin{itemize}
\item \emph{Infinite unfolding} happens when a call to a function
  \texttt{F} is encountered during partial evaluation of the same
  function \texttt{F}, indefinitely, just as in ordinary
  non-terminating recursion.

\item \emph{Infinite specialization} happens when partial evaluation
  attempts to create an infinite number of specialized versions of
  some function, such as \texttt{ADD(1,\#NA)}, \texttt{ADD(2,\#NA)},
  \texttt{ADD(3,\#NA)}, and so on.
\end{itemize}

\noindent
Moreover, we would like to avoid generating a finite but large number
of specialized versions of a function, when these turn out to be
nearly identical and offer no significant speed-up.  This particular
problem is discussed in
section~\ref{sec-partial-evaluation-termination-generalization}. 

Some of the problems we need to address are:

\begin{itemize}
\item When should we create further specializations of a sheet-defined
  function, while already in the process of creating one specialization?

\item More precisely, given a sheet-defined function and static values
  of some of its arguments, which of these arguments should actually
  be used when specializing the function?  Specializing with respect
  to all-\texttt{\#NA} arguments should be equivalent to not
  specializing the function at all, thus making \texttt{SPECIALIZE}
  idempotent.  This is convenient when both a higher-order library
  function and its caller attempt to specialize an argument closure.

\item When all arguments to a sheet-defined function, encountered
  during partial evaluation, are static, should we then attempt to
  fully evaluate the function or should we specialize it?
\end{itemize}

\subsubsection{Principle 1: Residualize all function calls}

When the result of a function would be non-static (because some
arguments are non-static or because the function body contains a call
to a volatile or external function), then we would lose little
information by not unfolding the call but reducing it to a call to a
residual function.  This decision does not affect the meaning or
termination properties of the partial evaluation process.

However, when the result of partially evaluating the function body
would be a concrete static value, then unfolding would propagate the
concrete value to the call context, thus enabling further computation,
whereas residualizing the call will lose the information held in that
static value.  

Nevertheless, we decide to always residualize and never unfold a
function call:

\begin{itemize}
\item Principle 1: The result of partially evaluating a function call
  is a residual function call.
\end{itemize}

\noindent
This ensures that the only source of non-termination during partial
evaluation is infinite specialization.

We residualize although in Funcalc it would be trivial to invoke the
standard evaluation machinery and then wrap the resulting Value as an
appropriate GCConst object
(figure~\ref{fig-funcalc-cgexpr-classdiagram}), whenever all arguments
to a function call are static.  However, using the standard evaluation
machinery during partial evaluation would be wrong, because it will
evaluate any volatile and external functions prematurely: even when
all needed arguments are static, the result may be a residual
expression instead of a value.  To see this, consider the function in
example~\ref{ex-sdf-expsample}.

\begin{example}\label{ex-sdf-expsample}
  Function \texttt{EXPSAMPLE} permits sampling
  from the exponential distribution:

{\small
\begin{verbatim}
  EXPSAMPLE(p,n) = IF(p<=0.0, ERR("P"), 
                      IF(RAND()<p, n, EXPSAMPLE(p,n+1)))
\end{verbatim}}

\noindent
Function \texttt{EXPSAMPLE(p,n)} either terminates immediately, with
probability \texttt{p}, or otherwise performs one more recursive call.
Thus \texttt{EXPSAMPLE(1,1)} will return 1 immediately;
and \texttt{EXPSAMPLE(0.5,1)} will return 1 with probability 0.5,
will return 2 with probability 0.25, will return 3 with probability
0.0125, and so on, that is, on average will return 2;
\texttt{EXPSAMPLE(p,1)} will on average return \texttt{1/p}, the mean
value of the exponential distribution with parameter \texttt{p}, when
$0 < \mathtt{p} \leq 1$.
\end{example}

\noindent
When both arguments to \texttt{EXPSAMPLE(p,n)} are static, the
arguments of the recursive call will be static too.  But since the
condition \texttt{RAND()<p} is volatile and will be residualized, the
IF-expression will be residualized too, and unfolding of the recursive
function call would go on indefinitely, in an attempt to construct an
infinite tree of residual conditionals.

To avoid such construction of infinite residual terms, we decide
never to unfold function calls.

\subsubsection{Principle 2: Generalizing under dynamic control}
\label{sec-partial-evaluation-generalizing}

In the \texttt{EXPSAMPLE} case, the decision never to unfold a
function call will avoid infinite unfolding, but might cause infinite
specialization instead, in an attempt to create specialized versions
for \texttt{n} being 1, 2, 3, and so on.

To avoid this, we need to \emph{generalize} one or more static
arguments, here \texttt{n}, by reclassifying them as dynamic, that is,
consider them to be \texttt{\#NA} in further specialization.  This
works as follows.

We say that an expression is under \emph{dynamic control} if some
conditional (\texttt{IF}, \texttt{CHOOSE}) with dynamic condition has
been encountered before the expression during the specialization
process.  Whether an expression appears under dynamic control can be
determined by passing a context argument along with the partial
evaluation environment \texttt{pEnv} in the recursive calls to the
partial evaluator.

If, in the process of specializing a sheet-defined function \texttt{F}
with respect to values \texttt{v1...vn}, we encounter a recursive call
to \texttt{F(...)}\ under dynamic control and with arguments
\texttt{w1...wn}, then we specialize \texttt{F} in the recursive call
only with respect to those static arguments \texttt{wj} that have the
same constant value \texttt{wj=vj} in both calls; the remaining static
\texttt{wj} are made dynamic, that is, considered \texttt{\#NA} when
specializing the called function.

Although this may seem draconically conservative, it will serve one large
class of specialization cases well, namely when some static
``configuration'' or ``problem'' parameters are passed to the function
initially, and passed on unchanged in all recursive calls.  Such
static parameters will be inlined (and possibly cause \texttt{IF} and
\texttt{CHOOSE} expressions to be reduced) but the part of the control
structure that depends also on dynamic parameters will be preserved.  The
draconic policy could be loosened a little by permitting specialization
with respect to literal constants given in the function (since there
are only finitely many of those), but it is unclear whether this is
worthwhile in general, and we currently do not do it.

To implement the above policy, we need a partial evaluation context
(an evaluation stack) that says which functions are currently being
specialized with respect to which constant arguments, and the value of
those arguments (so we can deal with mutually recursive functions),
and an indication whether the current expression (especially call) is
under dynamic control.  A partial evaluation context must tell us (1)
whether the expression being partially evaluated is under dynamic
control, so we can decide what to do with calls CGApply or CGSdfCall;
(2) which functions are currently being partially evaluated, so we can
recognize recursive calls; and (3) the arguments given to the functions
currently being partially evaluated.

Property (1) is a local property of a subexpression of a ComputeCell,
determined by the cell's evaluation condition and the conditions
enclosing that subexpression in the cell.  This notion of context
could therefore be represented by an argument IsDynamicControl passed
in as an argument of the partial evaluation process.  For an
evaluation condition it is initially false; for the expression in a
ComputeCell it is true iff its evaluation condition is dynamic.  For
CgIf and CGChoose it is determined as one might expect, by the first
argument.  Note the
difference from the partial evaluation environment \texttt{pEnv},
which grows monotonically while processing the list of ComputeCells
belonging to a given sheet-defined function.

   
Properties (2) and (3) are somewhat more global.  They too could be
represented by a parameter to the partial evaluator, but would need
much broader scope: not only the partial evaluation of a given
sheet-defined function, but a family of such functions.

To see the effect of the simple generalization technique, consider
these examples.

\begin{example}\label{ex-sdf-ackermann-A}
  Ackermann's function is sometimes used to illustrate partial
  evaluation of recursive functions \cite[section
  17.3]{Jones:1993:PartialEvaluation}. It may be defined like
  this:

{\small
\begin{verbatim}
  ackA(m,n) = IF(m=0, n+1, IF(n=0, ackA(m-1,1), ackA(m-1, ackA(m, n-1))))
\end{verbatim}}

\noindent
If we assume that \texttt{m} is static and equal to 2, and \texttt{n}
is dynamic, then the outer \texttt{IF} is static but the inner one is
dynamic.  Using the generalization strategy outlined above, we get the
following specialized function:

{\small
\begin{verbatim}
  ackA2(n) = IF(n=0, ackA(1,1), ackA(1, ackA2(n-1)))
\end{verbatim}}

\noindent
which basically specializes only with respect to the first value of
\texttt{m}, and only at the top-most call to \texttt{ackA} and
therefore misses significant optimization opportunities.  But to make
sure that the specialization of \texttt{ackA} with respect to static
first argument \texttt{m-1} will terminate, we really need to know
that the first argument is descending and bounded from below (as in
several static termination analyses).  This requires a somewhat
sophisticated static analysis, made even more complicated by the
dynamically typed spreadsheet formulas, so we prefer to avoid it.
\end{example}

\noindent
The next example shows that by rewriting the Ackermann function in a
slightly different style, we achieve much better specialization under
the same generalization scheme.

\begin{example}\label{ex-sdf-ackermann-B}
  Now define Ackermann's function like this, pushing the inner
  conditional inside the recursive call:

{\small
\begin{verbatim}
   ackB(m,n) = IF(m=0, n+1, ackB(m-1, IF(n=0, 1, ackB(m, n-1))))
\end{verbatim}}

\noindent
Then we get the following much better specialization for \texttt{m=2}
static and \texttt{n} dynamic:

{\small
\begin{verbatim}
   ackB2(n) = ackB1(IF(n=0, 1, ackB2(n-1)))
   ackB1(n) = ackB0(IF(n=0, 1, ackB1(n-1)))
   ackB0(n) = n+1
\end{verbatim}}

\noindent
This is just as we would like it, and similar to what one would get
from an off-line partial evaluator and a binding-time analysis, even
on the original source program in example~\ref{ex-sdf-ackermann-A}.
\end{example}

\subsubsection{Termination}
\label{sec-partial-evaluation-termination-generalization}

The simple principles of (1) never unfold, and (2) generalize
aggressively when under dynamic control, achieve fairly good
termination properties of the partial evaluator.  The downside is that
the partial evaluator is somewhat conservative, although it achieves
reasonably good results on examples such as those in
section~\ref{sec-partial-evaluation-examples}.

Termination is far from guaranteed, though, and the partial evaluator
is far from fool-proof.  For instance, specialization can still fail
to terminate due to infinite specialization.  To see this, consider
a function definition such as the ``counting-down factorial'' function
\cite[page 172]{Weise:1991:AutomaticOnline}:

{\small
\begin{verbatim}
   FACD(N) = IF(N=0, 1, N * FACD(N-1))
\end{verbatim}}

\noindent
and assume that we attempt to specialize \texttt{FACD(-1)}\@.  Since
there is no dynamic control, the specializer will not generalize the
static argument and will attempt to create an infinite number of
specialized functions corresponding to the calls \texttt{FACD(-1)},
\texttt{FACD(-2)}, \texttt{FACD(-3)}, and so on.  However, standard
evaluation of \texttt{FACD(-1)} would also fail to terminate.

If the call \texttt{FACD(-1)} were under dynamic control, the
generalization mechanism would ensure that the argument gets
generalized, and only one specialized version will be generated.  

Holst's ``poor man's generalization'' \cite{Holst:1988:PoorMans}
\cite[section 7.2.2]{Jones:1993:PartialEvaluation} is a generalization
technique that will generalize any static parameter that is not used
to evaluate a conditional such as \texttt{IF} or \texttt{CHOOSE}\@.
The reasoning is that such a parameter will have little effect on the
residual program's size and that knowing its value will not help
termination of specialization, but might still give rise to an
excessive number of residual functions.  Using poor man's
generalization will not help ensure termination, but could help keep
the number of specialized versions in check, avoiding fruitless and
costly code generation that gives little performance benefit.
However, it is comparatively difficult to implement.  We would need a
backwards data flow analysis to properly determine which parameters
may (or must) eventually determine a conditional.  Moreover, this
analysis must be interprocedural and the language is higher-order, so
this is not entirely straightforward, and we consider it too much
complexity for a modest gain.

More advanced generalization schemes and termination analyses, such as
homeomorphic embedding, seem hard to use in the spreadsheet context
because most data are numbers or arrays of numbers; although
tree-structured data are representable as nested arrays this is not
natural to the spreadsheet context.  Hence we currently do not use
flow analysis-based generalization, only the online generalization
mechanism described above.

A simple expedient would be to put an arbitrary limit on the number of
specializations created for any given function.  Since the original
function is available anyway, it can be used as a fallback when the
limit is reached.  See also section~\ref{sec-extensions-future-work}.

\subsection{Simplification of arithmetic expressions}

During partial evaluation it is natural to use mathematical identities
to simplify arithmetical expressions.  For instance, $e + 0$ may be
reduced to $e$.  The full list of reductions, implemented in class
CGArithmetics2, is shown in
figure~\ref{fig-specialization-arithmetic-simplification}.

Some ``obvious'' mathematical identities, such as reducing $e * 0$ to
$0$, do not in general preserve spreadsheet semantics because $e * 0$
will evaluate to an error if $e$ does, whereas $0$ will not.
Nevertheless we have implemented all the listed reductions.  Also, it
may seem wrong in general to replace $e \verb|^| 0$ and $1 \verb|^| e$
with $1$, but the \textsc{ieee754} floating-point standard
\cite[section 9.2.1]{IEEE:2008:FloatingPoint} does prescribe these
identities for all values of $e$, even NaN\@.

\begin{figure}[htbp]
  \centering
  \begin{tabular}{lcll}\hline\hline
    Original & & Simplified & Note \\\hline
    $0 + e$ & $\longrightarrow$ & $e$\\
    $e + 0$ & $\longrightarrow$ & $e$\\
    $e - 0$ & $\longrightarrow$ & $e$\\
    $0 - e$ & $\longrightarrow$ & $-e$\\
    $e * 0$ & $\longrightarrow$ & $0$ & (*)\\
    $0 * e$ & $\longrightarrow$ & $0$ & (*)\\
    $1 * e$ & $\longrightarrow$ & $e$\\
    $e * 1$ & $\longrightarrow$ & $e$\\
    $e / 1$ & $\longrightarrow$ & $e$\\
    $e \verb|^| 1$ & $\longrightarrow$ & $e$ \\
    $e \verb|^| 0$ & $\longrightarrow$ & $1$ & (**)\\
    $1 \verb|^| e$ & $\longrightarrow$ & $1$ & (**)\\\hline\hline
      \end{tabular}
      \caption{Arithmetic simplifications performed by partial
        evaluation.  Those marked (*) or (**) may not preserve
        spreadsheet semantics; those marked (**) do agree with the
        \textsc{ieee754} floating-point standard.}
  \label{fig-specialization-arithmetic-simplification}
\end{figure}

\section{Partial evaluation examples}
\label{sec-partial-evaluation-examples}

\begin{example}\label{ex-partial-evaluation-monthlen}
  Function \texttt{MONTHLEN(y,m)} computes the
  length of month \texttt{m} in year \texttt{y}, taking leap years into
  account:
  
{\small
\begin{verbatim}
   MONTHLEN(y,m) = 
     CHOOSE(m, 31, 
            28+OR(AND(NOT(MOD(y, 4)), MOD(y, 100)), NOT(MOD(y, 400))), 
            31, 30, 31, 30, 31, 31, 30, 31, 30, 31)
\end{verbatim}}

\noindent
Specializing this function to a fixed month \texttt{m} will either
leave only the leap year logic, eliminating the switch (when
\texttt{m} is 2), or eliminate both that logic and the switch (when
\texttt{m} is not 2).  Here is the residual function for
\texttt{MONTHLEN(\#NA,3)}:

{\small
\begin{verbatim}
   0000: ldc.r8     31
   0009: call       NumberValue.Make(Double)
   000e: ret        
\end{verbatim}}

\noindent
On the other hand, specializing \texttt{MONTHLEN} to a given year,
such as 2012, produces a function where all the logic concerning leap
years has been removed.  This is the bytecode for the specialization
of \texttt{MONTHLEN(2012,\#NA)}; note in line \texttt{007f} the result
29 of computing \texttt{28+1} at specialization time:

{\small
\begin{verbatim}
   0000: ldarg      V_0
   0004: call       Value.ToDoubleOrNan
   0009: stloc.0    
   000a: ldloc.0    
   000b: call       IsInfinity(Double)
   0010: brtrue     0150
   0015: ldloc.0    
   0016: call       IsNaN(Double)
   001b: brtrue     0150
   0020: ldloc.0    
   0021: conv.i4    
   0022: ldc.i4     1
   0027: sub        
   0028: switch     (006c, 007f, 0092, 00a5, 00b8, 00cb,   // CHOOSE
                     00de, 00f1, 0104, 0117, 012a, 013d)
   005d: ldc.i4     5
   0062: call       ErrorValue.FromIndex(Int32)
   0067: br         014b
   006c: ldc.r8     31
   0075: call       NumberValue.Make(Double)
   007a: br         014b
   007f: ldc.r8     29                                     // reduced from 28+...
   0088: call       NumberValue.Make(Double)
   008d: br         014b
   0092: ldc.r8     31
   00ae: call       NumberValue.Make(Double)
   ... and so on for April through November ...
   0138: br         014b
   013d: ldc.r8     31
   0146: call       NumberValue.Make(Double)
   014b: br         015a
   0150: ldc.i4     2
   0155: call       ErrorValue.FromIndex(Int32)
   015a: ret        
\end{verbatim}}

\end{example}

\begin{example}\label{ex-partial-evaluation-rept4}
  Function \texttt{REPT4(s,n)} from
  example~\ref{ex-sdf-rept4}, which computes string \texttt{s}
  concatenated with itself \texttt{n} times, can be specialized with
  respect to a given string \texttt{s} or with respect to a given
  number \texttt{n}.

  Specialization with respect to a given string, as in
  \texttt{REPT4("abc",\#NA)}, achieves nothing useful.  The bytecode
  for the resulting specialized function is nearly identical to that
  for the original \texttt{REPT4}\@.  Both are 123 bytecode
  instructions long (some of which implement evaluation conditions),
  and identical except that the specialized function loads the string
  \texttt{"ABC"} from a table of string values, whereas the original
  one takes it from the first function argument.

  Specialization with respect to a given \texttt{n}, as in
  \texttt{REPT4(\#NA,7)}, is much more interesting.  Since the second
  parameter \texttt{n} is static and determines all conditionals, that
  parameter and all tests on it will be eliminated.  The result is not
  one but four specialized functions, corresponding to the values of
  \texttt{n} encountered in the recursive calls, namely 7, 3, 1 and 0.

  This is \texttt{REPT4(\#NA,7)\#201} corresponding to \texttt{n}
  being 7:
  
{\small
\begin{verbatim}
   0000: ldsfld     SdfManager.sdfDelegates
   0005: ldc.i4     202
   000a: ldelem.ref 
   000b: castclass  System.Func`2[Value,Value]
   0010: ldarg      V_0                         // load arg s
   0014: call       Invoke                      // call #202(s) giving r
   0019: stloc.3    
   001a: ldarg      V_0
   001e: ldloc.3    
   001f: call       Function.ExcelConcat        // r & r 
   0024: ldloc.3    
   0025: call       Function.ExcelConcat        // (r & r) & s
   002a: ret        
\end{verbatim}}

\noindent
Function \texttt{\#201} above computes \texttt{s}$^7$ for any string
\texttt{s}, by calling function \texttt{\#202} to compute
\texttt{s}$^3$ and then concatenating the result \texttt{r} with
itself and with \texttt{s} to obtain \texttt{s}$^7$.

Function \texttt{\#202} corresponds to \texttt{n} being 3 and has
exactly the same structure.  It calls function \texttt{\#203} to
compute \texttt{s}$^1$ and then concatenates the result with itself
and with \texttt{s} to obtain \texttt{s}$^3$.  

Function \texttt{\#203} corresponds to \texttt{n} being 1 and has the
same structure as the previous two.  It calls function \texttt{\#204}
which computes \texttt{s}$^0$, that is, the empty string:

{\small
\begin{verbatim}
0000: ldsfld     TextValue.EMPTY
0005: ret                                       // return ""
\end{verbatim}}

\noindent
Whereas calling the original function \texttt{REPT4("abc",7)} takes
1200 ns/call, calling the specialized closure \texttt{REPT4(\#NA,7)}
on argument \texttt{"abc"} takes only 524 ns/call.  Further speedup
could be achieved by inlining the calls to the auxiliary specialized
functions.
\end{example}

\begin{example}\label{ex-partial-evaluation-add-multistage}
  To illustrate multistage specialization, consider the three-argument
  function \texttt{ADD3(x,y,z)}:
  
{\small
\begin{verbatim}
   ADD3(x,y,z) = x+y+z
\end{verbatim}}

\noindent
The original bytecode for the unspecialized \texttt{ADD3} is this:

{\small
\begin{verbatim}
   0000: ldarg      V_0
   0004: call       Value.ToDoubleOrNan         // Unwrap arg x
   0009: ldarg      V_1
   000d: call       Value.ToDoubleOrNan         // Unwrap arg y
   0012: add        
   0013: ldarg      V_2
   0017: call       Value.ToDoubleOrNan         // Unwrap arg z
   001c: add        
   001d: call       NumberValue.Make(Double)    // Wrap result
   0022: ret        
\end{verbatim}}

\noindent
The function \texttt{ADD3(11,\#NA,\#NA)\#20}, resulting from
specializing \texttt{ADD3} to its first argument having value 11, is
this two-argument function:

{\small
\begin{verbatim}
   0000: ldc.r8     11
   0009: ldarg      V_0
   000d: call       Value.ToDoubleOrNan         // Unwrap arg y
   0012: add        
   0013: ldarg      V_1
   0017: call       Value.ToDoubleOrNan         // Unwrap arg z
   001c: add        
   001d: call       NumberValue.Make(Double)    // Wrap result
   0022: ret        
\end{verbatim}}

\noindent
The function \texttt{ADD3(11,\#NA,\#NA)\#20(23,\#NA)\#21}, resulting
from further specializing that function to its first remaining
argument having value 23, is this one-argument function:

{\small
\begin{verbatim}
   0000: ldc.r8     34
   0009: ldarg      V_0
   000d: call       Value.ToDoubleOrNan         // Unwrap arg z
   0012: add        
   0013: call       NumberValue.Make(Double)    // Wrap result
   0018: ret        
\end{verbatim}}

\noindent
Finally, the function
\texttt{ADD3(11,\#NA,\#NA)\#20(23,\#NA)\#21(32)\#22}, resulting from
specializing the above function to its last remaining argument having
value 32, is this zero-argument function:

{\small
\begin{verbatim}
   0000: ldc.r8     66
   0009: call       NumberValue.Make(Double)    // Wrap 66 as result
   000e: ret        
\end{verbatim}}

\noindent
The execution times of the above four functions are the following: 59,
45, 39, 35 ns/call.  Much of this cost, roughly 23 ns/call, arises not
from parameter passing, parameter unwrapping, or the addition
operations, but from the final wrapping of a floating-point result as
a NumberValue object.
\end{example}

\begin{example}\label{ex-partial-evaluation-expsample}
  Specializing \texttt{EXPSAMPLE(0.15,1)} from
  example~\ref{ex-sdf-expsample} with respect to static values of
  \emph{both} its arguments does not produce a number, because the
  original function involves the volatile \texttt{RAND()} function.
  Instead we get this argumentless residual function (\texttt{\#25}): 
  
{\small
\begin{verbatim}
   0000: call       ExcelRand()
   0005: ldc.r8     0.15
   000e: bge        001d                        // if RAND() < 0.15
   0013: ldsfld     NumberValue.ONE             //   return 1
   0018: br         0043
   001d: ldsfld     SdfManager.sdfDelegates     // else
   0022: ldc.i4     26
   0027: ldelem.ref 
   0028: castclass  System.Func`2[Value,Value]
   002d: ldc.r8     2
   0036: call       NumberValue.Make(Double)
   003b: tail.                                  //   tail call
   003d: call       Invoke(Value)               //   call #26 on 2
   0042: ret        
   0043: ret        
\end{verbatim}}

\noindent
The original \texttt{EXPSAMPLE} function from
example~\ref{ex-sdf-expsample} calls itself recursively with arguments
\texttt{(0.15,2)}, where the static argument \texttt{2} differs from
the previous value \texttt{1}.  Since the recursive call is under
dynamic control (section~\ref{sec-partial-evaluation-function-calls})
the second argument gets generalized to \texttt{\#NA}, so the
recursive call becomes a call to the specialization of
\texttt{EXPSAMPLE(0.15,\#NA)}\@.  The call appears above as a call to
function \texttt{\#26} which has this bytecode:

{\small
\begin{verbatim}
   0000: call       ExcelRand()
   0005: ldc.r8     0.15
   000e: bge        001c                        // if RAND() < 0.15
   0013: ldarg      V_0                         //   return n
   0017: br         004c                        
   001c: ldsfld     SdfManager.sdfDelegates     // else
   0021: ldc.i4     26
   0026: ldelem.ref 
   0027: castclass  System.Func`2[Value,Value]
   002c: ldarg      V_0
   0030: call       Value.ToDoubleOrNan(Value)
   0035: ldc.r8     1
   003e: add        
   003f: call       NumberValue.Make(Double)
   0044: tail.                                  //   tail call
   0046: call       Invoke(Value)               //   call #26 on (n+1)
   004b: ret        
   004c: ret        
\end{verbatim}}

\noindent
This residual function calls itself recursively, as function
\texttt{\#26}.  This result makes perfect sense, because different
calls to the argumentless result of
\texttt{SPECIALIZE(CLOSURE("EXPSAMPLE", 0.15, 1))} will produce
different samples from the exponential distribution with parameter
0.15.
\end{example}

\section{Other features}
\label{sec-other-features}

A companion paper \cite{Sestoft:2013:SheetDefined} describes
\emph{evaluation conditions}, a form of strictness analysis in logical
form needed when compiling sheet-defined functions that call
themselves recursively.  That paper also describes the experience of
implementing most of Excel's built-in financial functions as
sheet-defined functions.  

To facilitate reusing external code, such as linear algebra libraries,
we provide a foreign-function interface through the built-in function
\texttt{EXTERN}\@.  The implementation uses reflection, caching and
runtime code generation to obtain high performance and avoid
conversion of values on the interface between spreadsheet formulas and
the underlying CLI runtime system.

A built-in function \texttt{BENCHMARK(fv,count)} is provided to
measure the time (in ns) to call to 0-arity closure \texttt{fv},
measured as the average of \texttt{count} calls.  See
also figure~\ref{fig-clo-spec-apply-bench}.

\section{Related work}

Peyton-Jones, Blackwell and Burnett proposed
\cite{PeytonJones:2003:AUserCentred} that user-defined functions
should be definable as so-called \emph{function sheets} using ordinary
spreadsheet formulas, but their ideas were not implemented.  Similar
ideas are found in Nuñez's spreadsheet system ViSSh \cite[section
5.2.2]{Nunez:2000:AnExtended}.  Our concept of sheet-defined function
is strongly inspired by Peyton-Jones et al., but extends
expressiveness by permitting recursive and higher-order functions.
Resolver One \cite{ResolverSystems::ResolverOne} is a commercial
Python-based spreadsheet program with a feature called
\texttt{RUNWORKBOOK} that allows a workbook to be invoked as a
function, similar to a sheet-defined function at a coarser
granularity.

Unlike our work, those works do not emphasize the performance of
user-defined functions.  We believe that good performance is crucial
to the prospect of replacing built-in functions by libraries of
user-definable functions, and therefore built this prototype to see
what performance can be achieved by a simple and fast compiler
generating bytecode. 

Online partial evaluation was investigated in depth by Ruf, Weise and
others in the context of the Fuse specializer for Scheme, which is
dynamically typed and higher-order, just like our spreadsheet function
language
\cite{Ruf:1993:TopicsIn,Ruf:1992:OpportunitiesFor,Weise:1991:AutomaticOnline}.
One of the first notable online partial evaluators for a dynamically
typed language, considerably predating Fuse, was Redfun by Haraldsson,
Sandewall and others \cite{Haraldsson:1977:AProgram}.

Generalization as a means to ensuring termination of specialization
was central already to Turchin's early Refal work
\cite{Turchin:1988:TheAlgorithm}, and its importance was recognized
also in later work on off-line
\cite{Bondorf:1991:AutomaticAutoprojection,Jones:1989:Mix} and on-line
\cite{Weise:1991:AutomaticOnline} partial evaluation.  Our
generalization principle from
section~\ref{sec-partial-evaluation-generalizing} seems to be a simple
instance of Weise et al.'s type-based generalization, or
``abstraction''.  In our case, their notion of ``type'' or symbolic
specialization time value \cite{Weise:1991:AutomaticOnline} is simply
the expression representing it.  This is consistent with our
generalization of two equal values (represented by the same constant)
to that constant, and generalization of two distinct values to a
completely unknown value, preserving no (partial) type information,
unlike Fuse\@.

\section{Desirable extensions and future work}
\label{sec-extensions-future-work}

It would be desirable to have a better generalization strategy,
especially one whose termination properties are well understood.
Also, the generation of useless specializations should be prevented.
As a less desirable alternative, there could be mechanisms to
interactively control and tame excess generation of specialized
functions.  For instance, there might be a way to turn off
specialization once a certain number of specialized functions have been
generated, or to manually interrupt it. 

Likewise, one may ask whether the resulting specialized function is
correct.  This clearly depends on the expected semantics of
sheet-defined functions, which in turn depends on the expected
semantics of spreadsheet computations.  Although formalized nowhere,
to our knowledge, this semantics is mostly obvious, with the exception
of (1) error values and their propagation, and (2) the meaning of
volatile functions such as \texttt{RAND()} and \texttt{NOW()}\@.  We
believe our treatment of these, described in
section~\ref{sec-partial-evaluation-cgexpr-terms}, is sensible, but
would like to have a formal semantics with which to underpin this
claim.

In general-purpose languages it is difficult to estimate the cost
(specialization time, residual code size) and benefit (speedup) of
partial evaluation.  In the context of spreadsheets, where
``iteration'' is often implemented simply by making a sufficient
number of copies of a computation, it may be easier to estimate
whether specialization is worth-while.  For instance, it may be
evident from a worksheet that the function obtained by
\texttt{CLOSURE("name",a1,...,aM)} will be called from, say, 500 cells
due to replicated formulas.  A support graph \cite[chapter
4]{Sestoft:2012:SpreadsheetTechnology} provides a estimate of this
very cheaply; just count the number of cells directly supported by the
cell that evaluates the \texttt{CLOSURE}-expression.  Such estimates
are far harder in general functional and procedural languages.  Of
course, the more sophisticated the spreadsheet model is, the harder it
may be to obtain good estimates.  If users replace explicit formula
replication with recursive functions, then the advantages relative to
traditional languages are diminished.
 
It seems that partial evaluation can be especially beneficial in
connection with code generation for general-purpose graphics
processors (GPGPU)\@.  A graphics processor can efficiently run many
instances of the same straight-line numeric code in parallel, but it
is poorly equipped for executing branching code, such as that
resulting from the translation of \texttt{IF(\ldots)} and
\texttt{CHOOSE(\ldots)} in spreadsheet formulas.  So whereas partial
evaluation, with inlining of constants and early evaluation of
conditionals, offers modest speed-ups on a general CPU, it may offer
more impressive speedups when the code is to be executed on graphics
processors.

\section{Conclusion}

We have outlined an implementation of sheet-defined functions,
user-defined functions wholly based on standard spreadsheet concepts,
as suggested by Peyton-Jones and others
\cite{PeytonJones:2003:AUserCentred}.  We have shown that very good
performance can be achieved using relatively simple compiler
technology and runtime bytecode generation, exploiting the
just-in-time native code generation of the Microsoft .NET runtime
system.

The particular contribution of this paper is to show that runtime
partial evaluation, or automatic specialization, of sheet-defined
functions is feasible and can achieve further performance gains.  We
believe that the interactive and side-effect free spreadsheet setting
is a plausible context for practical use of partial evaluation.

\paragraph{Acknowledgements} Thanks to the participants at Meta 2010
in Pereslavl-Zalessky for comments, and to the anonymous referees for
their many constructive suggestions.

\bibliographystyle{eptcs}
\bibliography{spreadsheet,ownpapers}
\end{document}